\documentclass[preprint,aps,prb]{revtex4-1}
\usepackage{graphicx}
\usepackage{amssymb}
\usepackage{amsfonts}
\usepackage{amsmath}
\usepackage{amsbsy}
\usepackage{natbib}
\usepackage{comment}
\usepackage{hyperref}
\usepackage{color}
\usepackage{float}

\usepackage[utf8]{inputenc}
\usepackage[T1]{fontenc}
\usepackage{amsthm}
\usepackage{scalerel}

\usepackage{lmodern}
\usepackage{graphicx}
\usepackage{mathtools}

\newcommand{\eu}{\ensuremath{\mathrm{e}}}	
\newcommand{\iu}{\ensuremath{\mathrm{i}}}	
\newcommand{\du}{\ensuremath{\mathrm{d}}}	

\begin{document}

\title{ Magneto-Electric Response of Quantum Structures    Driven by Optical Vector  Beams}

\author{Jonas W\"{a}tzel, C. M. Granados--Castro, and Jamal Berakdar}
\affiliation{Institute for Physics, Martin-Luther-University Halle-Wittenberg, 06099 Halle, Germany}
%

\begin{abstract}
Key advances  in the generation and shaping of spatially
structured photonic fields both in the near and far field  render possible the control of  the
duration, the phase, and the polarization  state of the field distributions. For instance, optical vortices having a structured phase
are nowadays routinely generated and exploited for a range of applications.
 While the light-matter interaction
with optical vortices is meanwhile well studied, the distinctive features  of the
interaction of quantum matter with \textit{vector beams}, meaning fields with spatially
inhomogeneous polarization states, are still to be explored in full detail, which
is done here. We analyze the response of atomic and low dimensional quantum structures  to irradiation
with radially or azimuthally polarized cylindrical vector
beams. Striking
differences to vortex beams are found:  Radially polarized vector
beams drive radially \textit{breathing charge-density oscillations} via electric-type
quantum transitions. Azimuthally polarized vector beams do not affect the charge at all
but trigger, via a magnetic vector potential a \emph{dynamic Aharonov--Bohm
effect}, meaning a vector-potential driven oscillating magnetic moment. In contrast to vortex beams, no
unidirectional currents are generated. Atoms driven by a radially polarized vector
beam exhibit angular momentum conserving quadrupole transitions tunable by a
static magnetic field, while when excited with azimuthally polarized beam
different final-state magnetic sublevels can be accessed.
\end{abstract}

\date{\today}
\maketitle

\section{Introduction}

Spatio-temporally modulated electromagnetic (EM) fields in general, and laser
fields in particular, have been the driving force for numerous discovery in science
as in femto-chemistry and attosecond physics both relying on the controlled
temporal shaping of laser fields~\cite{brabec2000intense, petek1997femtosecond}.
Spatially structured EM fields, which are  in the focus  of research currently,
have also proved instrumental for a wide range of applications such as particle
trapping~\cite{ng2010theory}, high-resolution lithography~\cite{hao2010phase, %
	dorn2003sharper, kang2012twisted, bauer2014nanointerferometric}, quantum memories~%
\cite{parigi2015storage}, optical communication~\cite{d2012complete, %
	vallone2014free}, classical entanglement~\cite{gabriel2011entangling}, and as a
magnetic nanoprobe for enhancing the  near field magnetic component~\cite{Guclu2016}.

Prominent examples of structured EM fields are orbital-angular momentum carrying
(OAM) vortex beams and vector beams (VBs). OAM beams possess an inhomogeneous
azimuthal phase distribution and a homogeneous polarization. For VBs the spatial
distributions of both the phase and the polarization in the plane perpendicular
to the propagation of the EM wave are inhomogeneous. The spatial structuring
brings about several advantages. For instance, a radially polarized VB allows for
a sharper focusing. It may also have a strong centered longitudinal field~\cite{%
	dorn2003sharper}, offering a tool for investigating new aspects of light-matter
interaction, as detailed below. On the other hand, an azimuthally polarized VB
has a smaller spot size than a radially polarized VB~\cite{hao2010phase} and
interact with quantum matter in fundamentally different manner, as shown here.
Phase modulated beams carrying OAM \cite{BLIOKH20151} serve further purposes. For instance, such
beams were used to study otherwise inaccessible angular momentum state of atoms~%
\cite{schmiegelow2016transfer} and to generate unidirectional steady-state charge
currents in molecular matter or in nanostructures ~\cite{Watzel2016, %
	watzel2016optical} pointing so to qualitatively new routes in optomagnetism.

Theoretically, key quantities for understanding the fundamental of the
interaction of structured fields with matter are the associated EM vector $\pmb{A}%
(\pmb{r},t)$ and scalar $\Phi(\pmb{r},t)$ potentials that couple, respectively to
the sample's currents and charge densities. Homogeneous optical EM fields
irradiating a quantum object (with a charge localization below the EM-field
wavelength) induce mainly electric-dipole transitions in the sample and to a
much smaller degree magnetic-dipole transitions. At moderate intensities, the
ratio of the magnetic dipole to the electric dipole absorption rate is proportional
to the ratio of the magnetic to the electric field strengths $\left|\pmb{H}%
\right|^2/ \left|\pmb{E}\right|^2$~\cite{Zurita-Sanchez2002}. Therefore, tailored
laser beams with engineered magnetic to electric-field ratio may boost the magnetic
transitions. For instance, this can be accomplished in the near-field of an object
with a small circular aperture~\cite{Hanewinkel1997}. For the nm apertures
experimentally feasible so far, the magnetic transitions enhancement is negligibly
small, however~\cite{Hanewinkel1997}. In this context, cylindrical VBs with
azimuthal or radial polarization offer an interesting alternative. For azimuthally
polarized VBs the magnetic to electric field ratio is substantial: one can show
that $\left|\pmb{H}\right|/\left|\pmb{E}\right|=1/\eta_0$ on the beam axis where
$\eta_0$ is the free-space impedance~\cite{veysi2015vortex, Zurita-Sanchez2002}.
The VBs we will be dealing with can be experimentally realized by the coherent
interference of two TEM$_{01}$ laser modes which are orthogonally polarized~\cite{%
	Oron2000}. Other techniques involve interferometry~\cite{Tidwell1990}, holograms~%
\cite{Churin1993}, liquid crystal polarizer~\cite{Stalder1996}, spatial light
modulators~\cite{Tripathi2012} and multi-elliptical core fibers~\cite{Milione2011}.
Planar fabrication technologies in connection with flat optics devices could also
produce cylindrical VBs~\cite{Memarzadeh2011, Yu2011, Yu2012}. A further approach
relies on the conversion of circularly polarized light into radially or azimuthally
VBs (in the far-infrared~\cite{Bomzon2002} and visible range~\cite{Beresna2011})
by space-variant gratings. A method involving an inhomogeneous half-wave plate
metasurface to generate VBs was also demonstrated~\cite{Yi2014, Liu2014a} where
the efficiency was increased when employing suitable  metamaterials~\cite{%
	veysi2015vortex}.

VBs possess a pronounced longitudinal component that can be employed for Raman-%
spectroscopy~\cite{saito2008z}, material-processing~\cite{Wang2008,Meier2007} or
tweezers for metallic particles~\cite{zhan2004trapping}. For OAM carrying beams,
the longitudinal component may serve for studying subband states in a quantum
well~\cite{sbierski2013} and hole-states in quantum dots~\cite{quinteiro2014light}.
The various facets of the  interactions of VBs with quantum matter will be
addressed in this present work. As a demonstration of the formal theory, we will
study the nature of bound-bound and bound-continuum transitions caused by VBs
when interacting with quantum systems such as nanostructures or atoms. As
demonstrated here the interaction of such fields is not only fundamentally
different from non-structured fields but also, radial VBs interact with matter in
a qualitatively different way as azimuthal VBs do, and the employment of both offer qualitatively new
opportunities for accessing the magneto-electric response of quantum matter at moderate intensities.


\section{\label{sec:sec1}Light-Matter Interaction with Cylindrical Vector Beams}

A cylindrical vector beam may be composed from two counter-rotating circularly
polarized optical vortex beams. The most prominent feature of such a vortex EM
beam is the azimuthal phase structure  described by  $\exp(\iu m_{\rm OAM}\varphi)$
where $\varphi$ is the azimuthal angle in the $xy$ plane~\cite{Allen1992orbital, %
	Yao2011}, transverse to the propagation direction (which sets the $z$-direction,
the radial distance we denote by $\rho$). The parameter $m_{\rm OAM}$ is the vortex
topological charge that  determines the amount of the carried orbital angular momentum
(OAM),  and can potentially  be transferred to a sample~\cite{Quinteiro2011,quinteiro2014light, Watzel2016}. The wave vector along $z$ is $q_{\parallel}$.
Optical vortices have a phase singularity at $\rho=0$ and thus a vanishing
intensity at this point. Generally, the transverse spatial distribution is
characterized by the function $f_{m_{\rm OAM}}(\rho)$ which can be of a
Laguerre--Gaussian type~\cite{Allen1992orbital}, Hermite--Gaussian~\cite{Novotny1998}
type or Bessel type~\cite{Garces-Chavez2002} with the main difference being the
radial intensity localization. For nano-scale objects centered in the vicinity of
the optical axis, the different radial distributions of diffraction limited vortex
beams have similar influence (due to the vast difference between electronic and
optical wave lengths). The change in this behavior with increasing  ${\rm OAM}$
can be inferred from the fact that $f_{m_{\rm OAM}}(\rho) \sim \rho^{\left|m_{%
		\rm OAM}\right|}$ for $\rho\rightarrow0$. As an example, we concentrate on  Bessel
beams which are exact solution of the Helmholtz equation~\cite{Garces-Chavez2002},
meaning that our theoretical considerations are beyond the paraxial approximation.
Bessel beams are non-diffracting beam solutions with the radial profiles being
independent of the propagation direction $z$ and the associated electromagnetic
vector potential is a solenoidal vector field: $\nabla\cdot\pmb{A}(\pmb{r},t)\equiv0$.
For the electromagnetic field components follows ($\pmb{r}=\left\{\rho,\varphi,%
z\right\}$ and $\Re$ means real part)
\begin{equation}
\pmb{A}(\pmb{r},t) = \Re\left\{\eu^{\iu(q_{\parallel}z-\omega t)}\left[\hat{e}_\sigma
J_{m_{\rm OAM}}(q_{\perp}\rho)\eu^{\iu m_{\rm OAM}\varphi}
\vphantom{\frac{q_{\perp}}{q_{\parallel}}}
- \iu\sigma\hat{e}_z\frac{q_{\perp}}{q_{\parallel}}J_{m_{\rm OAM}+\sigma}(q_{\perp}\rho)
\eu^{\iu(m_{\rm OAM}+\sigma)\varphi}\right]\right\}
\end{equation}
where $A_0$ is the vector potential amplitude and $\omega$ is the light frequency.
The longitudinal and radial wave vectors satisfy the relation $q_{\parallel}^2 +%
q_{\perp}^2 =(\omega /c)^2$. The functions $J_n(x)$ are the Bessel functions of
$n$th order while the polarization state is characterized by $\hat{e}_{\sigma}=%
\eu^{\iu\sigma\varphi}\left(\hat{e}_\rho + \iu\sigma\hat{e}_\varphi\right)$, with
$\sigma=\pm1$. The ratio $q_{\perp}/q_{\parallel}=:\tan\alpha$. Consequently, the
angle $\alpha$ characterizes the spatial extent of the intensity profile. A large
transverse wave vector means a tighter focusing. Since Bessel beams satisfy also
the Coulomb gauge, the electric field  reads $\pmb{E}(\pmb{r},t)=-\partial_t\pmb{A}%
(\pmb{r},t)$ while the magnetic field is given by $\pmb{B}(\pmb{r},t)=\pmb{\nabla}%
\times\pmb{A}(\pmb{r},t)$.

An  azimuthally polarized cylindrical VB (we refer to as AVB) can be expressed as a linear
combination of two optical vortices with $\{m_{\rm OAM}=+1, \sigma=-1\}$ and
$\{m_{\rm OAM}=-1, \sigma=+1\}$, namely
\begin{equation}
\pmb{A}_{\rm AVB}(\pmb{r},t)=A_0J_1(q_{\perp}\rho)\sin(q_{\parallel}z-\omega t)\hat{e}_{\varphi}.
\label{eq:A_AVB}
\end{equation}
A  radially polarized cylindrical  VB (denoted by  RVB) is expressible as the difference of the two optical
vortices
\begin{equation}
\pmb{A}_{\rm RVB}(\pmb{r},t)=A_0\left[-J_1(q_{\perp}\rho)\cos(q_{\parallel}z-\omega t)\hat{e}_{\rho}
\vphantom{\frac{q_{\perp}}{q_{\parallel}}}
+ \frac{q_{\perp}}{q_{\parallel}}J_0(q_{\perp}\rho)\sin(q_{\parallel}z-\omega t)\hat{e}_{z}\right].
\label{eq:A_RVB}
\end{equation}
A hallmark of  AVB and RVB  is the vanishing of the azimuthal-plane component of
the field at $\rho=0$. Moreover, AVB and RVB possess a non-vanishing longitudinal
component: beams with the azimuthal polarization have a magnetic component at the
origin while the longitudinal component of the radially polarized light mode (RVB)
is electric. The explicit electric and magnetic fields for both vector beam
classes can be found in the first section of the Supporting Information (SI).

We find that for both VBs the minimal coupling to matter is still viable leading
to the general interaction  operator $\hat{H}_{\rm int}$ with a collection of
charge carriers with  effective mass $m_e^*$ and charge $-e$
\begin{equation}
\begin{split}
\hat{H}_{\rm int,tot} &=\sum_i \hat{H}_{\rm int,i}, \\
\hat{H}_{\rm int,i} &=
-\frac{e}{2m_e^*} \left[\hat{\pmb{p}_i}\cdot\pmb{A}(\pmb{r}_i,t)
+ \pmb{A}(\pmb{r}_i,t)\cdot\hat{\pmb{p}_i}\right]
+ e\Phi(\pmb{r}_i,t)
\end{split}
\label{eq:Hint}
\end{equation}
$\hat{\pmb{p}_i}$ is the linear momentum operator of particle $i$  at the position
$\pmb{r}_i$ (for moderate intensities we may suppress the term $\pmb{A}^2(\pmb{r}_i,t)$).
%
%
It is instructive to exploit the gauge invariance of observables and go over to
the potentials (for brevity index $i$ is suppressed)
\begin{equation}\label{eq:ap}
\pmb{A}'(\pmb{r},t)=-\pmb{r}\times\int_0^1 \du\lambda\,\lambda\pmb{B}(\lambda\pmb{r},t)
\end{equation}
and
\begin{equation}\label{eq:fp}
\Phi'(\pmb{r},t)=-\pmb{r}\cdot\int_0^1 \du\lambda\,\pmb{E}(\lambda\pmb{r},t)
\end{equation}
The choice is referred to the Poincar\'{e} gauge or generally the multipole
gauge~\cite{Tannoudji1989, Tannoudji1992}.
Note that in this gauge $\pmb{r}\cdot\pmb{A}'(\pmb{r},t)\equiv0$ and $\pmb{B}(%
\pmb{r},t)=\pmb{\nabla}\times\pmb{A}'(\pmb{r},t)$ while $\pmb{E}(\pmb{r},t)=-%
\partial_t\pmb{A}'(\pmb{r},t) -\pmb{\nabla}\Phi'(\pmb{r},t)$. With eqs~\eqref{eq:ap}
and \eqref{eq:fp}, the light-matter interaction can be expressed as a sum of a pure
electric and magnetic contributions $\hat{H}_{\rm int}=\hat{H}_{\rm el}+ \hat{H}_{%
	\rm magn}$, where
\begin{equation}
\hat{H}_{\rm el}(t)=e\Phi'(\pmb{r},t)=e\,\pmb{r}\cdot\pmb{E}'(\pmb{r},t)
\label{eq:Helec}
\end{equation}
and $\pmb{E}'(\pmb{r},t)=-\int_0^1
\du\lambda\pmb{E}(\lambda\pmb{r},t)$. The magnetic part reads~\cite{Quinteiro2017}
\begin{equation}
\hat{H}_{\rm magn}(t)=2\pmb{B}'(\pmb{r},t)\cdot\hat{\pmb{m}}_B
\label{eq:Hmagn}
\end{equation}
with the field $\pmb{B}'(\pmb{r},t)=-\int_0^1 \du\lambda\,\lambda\pmb{B}(\lambda%
\pmb{r},t)$ and the magnetic moment operator $\hat{\pmb{m}}_B=(e/2m_0)\,\pmb{r}%
\times\hat{\pmb{p}}$  (for more details, see SI).
For a homogeneous field $\pmb{B}'(\pmb{r},t) = -\frac{1}{2}\pmb{B}(t)$ we obtain
the well-known dipolar magnetic interaction $\hat{H}_{\rm magn}(t) = -\hat{%
	\pmb{m}}_B\cdot\pmb{B}(t)$. Considering a spin-active system with a spin-dependent
field-free Hamiltonian $\hat{H}_0 $ such as (with $\hat{\pmb{\sigma}}$ being a
vector of Pauli matrices)
\begin{equation}
\hat{H}_{0}=\frac{\hat{\pmb{p}}^2}{2m_e^*} +
\frac{\alpha_R}{\hbar}\left[\hat{\pmb{\sigma}}\times\hat{\pmb{p}}\right]_z + V(\mathbf{r})
\label{eq:HGS}
\end{equation}
where  $V(\mathbf {r})$ is a scalar potential and $\alpha_R$ is a (Rashba)
spin-orbital interaction (SOI) strength, we find the following expression upon
applying the external VBs
\begin{equation}
\hat{H}=\hat{H}_0 + \hat{H}_{\rm int} (t)+ \hat{H}_{\rm SOI} (t)+
\hat{H}_{\rm z}(t)
\end{equation}
The  field-induced spin-orbital interaction $\hat{H}_{\rm SOI}(t)=-\frac{e%
	\alpha_R}{\hbar}\left[\hat{\pmb{\sigma}}\times\pmb{A}(
\pmb{r},t) \right]_z$ transforms in the Poincar\'{e} gauge to
\begin{equation}
\begin{split}
\hat{H}_{\rm SOI}(t)&=-\frac{e\alpha_R}{\hbar}\left[\hat{\pmb{\sigma}}\times%
\left(\pmb{r}\times\pmb{B}'(\pmb{r},t)\right)\right]_z
\end{split}
\label{eq:SOI_Bz}
\end{equation}
The field-induced Zeeman coupling reads
\begin{equation}
\hat{H}_{\rm z}(t)=-\frac{1}{2}\mu_B g_s\,\hat{\pmb{\sigma}}\cdot\pmb{B}(\pmb{r},t)
\end{equation}
where $\mu_B$ is the Bohr magneton and $g_s$ is the anomalous gyromagnetic ratio.
In the  static limit we recover the usual Zeeman coupling  lifting the  spin
degeneracy ~\cite{Fustaglia2004, Molnar2005, Sheng2006,Foldi2006, Zhu2008}.

\section{\label{sec:sec2} Spin-active Quantum Ring Structures}

For numerical demonstrations we consider quantum rings. Physical systems are, for
example, molecular macrocycles or rotaxane structures~\cite{iyoda2011conjugated, %
	choi2003macrocyclic, stoddart2017mechanically}. Here, we inspect
an appropriately doped quantum ring etched in a semi-conductor-based two-dimensional
electron gas. The conduction band charge carriers are tightly confined in the
direction normal to the ring plane by the  potential $U(z)$. In the ring plane the radially
symmetric potential $V(\rho)$ defines the ring. The independent charge carriers
are free to move in the  azimuthal direction $\hat{e}_\varphi$. The (spin-degenerate)
single particle states are represented by the wave functions $\Psi_{n,m,k}(\rho,%
\varphi,z)=\frac{1}{\sqrt{2\pi}} \rho^{-1/2}R_{nm}(\rho)\eu^{\iu m\varphi}Z_{k}(z)$
with the normalization $\int\du\rho\,R_{nm}(\rho)R_{n'm}(\rho)=\delta_{n,n'}$ and
$\int\du z\,Z_{k}(z)Z_{k'}(z)=\delta_{k,k'}$. The particles number  and $U(z)$
are chosen such that  only the lowest subband $k=0$ is occupied. This can be achieved in a semi-conductor based
structure by an appropriate gating.  Henceforth, we
omit therefore the index $k=0$ for brevity and trace out the $z$-dependence.
Furthermore, we checked that the driving field amplitude and its frequency do not
cause any transitions to subbands with $k\neq0$.

The time-independent single-particle Hamiltonian including SOI (\ref{eq:HGS})
has been already discussed extensively in several works~\cite{Fustaglia2004, %
	Molnar2005, Sheng2006, Foldi2006, Zhu2008}, albeit for homogeneous EM fields.
Considering intraband transition in the lowest radial subband $n=0$, the angular-%
dependent spin resolved single-particle wave functions are
\begin{equation}
\Psi_{m}^s=N_n \eu^{\iu(m+1/2)\varphi}\nu^s(\gamma,\varphi)
\end{equation}
where $s$ and $m$ denote the spin and integer angular quantum numbers and $N_n$
stands for the normalization. The spinors
\begin{equation}
\nu^s(\gamma,\varphi)=\left(a^s \eu^{-\iu\varphi/2},b^s \eu^{\iu\varphi/2}\right)^T
\end{equation}
are defined in the local frame with
\begin{equation}
a^\uparrow=\cos(\gamma/2),\,b^\uparrow=\sin(\gamma/2)
\end{equation}
and
\begin{equation}
a^\downarrow=-\sin(\gamma/2),\,b^\downarrow=\cos(\gamma/2)
\end{equation}
The angle $\gamma$ defines the direction of the spin relative to $\hat{e}_z$ with
a value set  by SOI strength: $\tan(\gamma)=-\omega_R/\omega_0$ where $\hbar%
\omega_R= 2\alpha_R/\rho_0$ and $\hbar\omega_0=\hbar^2/(m^{*}_e\rho_0^2)$, is the
inherent energy scale of a ring with a radius $\rho_0$. The local spin orientations
are inferred from the relations
\begin{equation}
\begin{split}
S_{\uparrow}(\pmb{r})= &\frac{\hbar}{2}\left[ \sin(\gamma)\cos(\varphi)\hat{e}_x
+ \sin(\gamma)\sin(\varphi)\hat{e}_y  + \cos(\gamma)\hat{e}_z\right]
\end{split}
\label{eq:orientation_SU}
\end{equation}
for the spin-up states, while the spin-down states are characterized by
\begin{equation}
\begin{split}
S_{\downarrow}(\pmb{r})=\frac{\hbar}{2}\left[ \sin(\pi-\gamma)\cos(\pi+\varphi)\hat{e}_x
+ \sin(\pi-\gamma)\sin(\pi+\varphi)\hat{e}_y
+ \cos(\pi-\gamma)\hat{e}_z\right].
\end{split}
\label{eq:orientation_SD}
\end{equation}
The associated eigenenergies are given by
\begin{equation}
E_{m}^s=\frac{\hbar\omega_0}{2}\left[ (m-x_s)^2-\frac{Q_R^2}{4} \right]
\label{eq:energy}
\end{equation}
where $x_s=-(1-sw)/2$ and $w=\sqrt{1+Q_R^2}=1/\cos(\gamma)$. Furthermore, $s=\pm1$
stand for up and down spin states. We emphasize that, hereafter, the terms up and
down (labeled, respectively, $\uparrow$ and $\downarrow$) refer to directions in
the local frame $\{\gamma,\varphi\}$ (cf. eqs~\eqref{eq:orientation_SU} and
\eqref{eq:orientation_SD}). The two characteristic spin bands are separated from
each other by $w$ which, in return, depends on the strength of the spin-orbit
coupling $\alpha_R$.

\begin{figure}[t]
	\includegraphics[width=0.95\columnwidth]{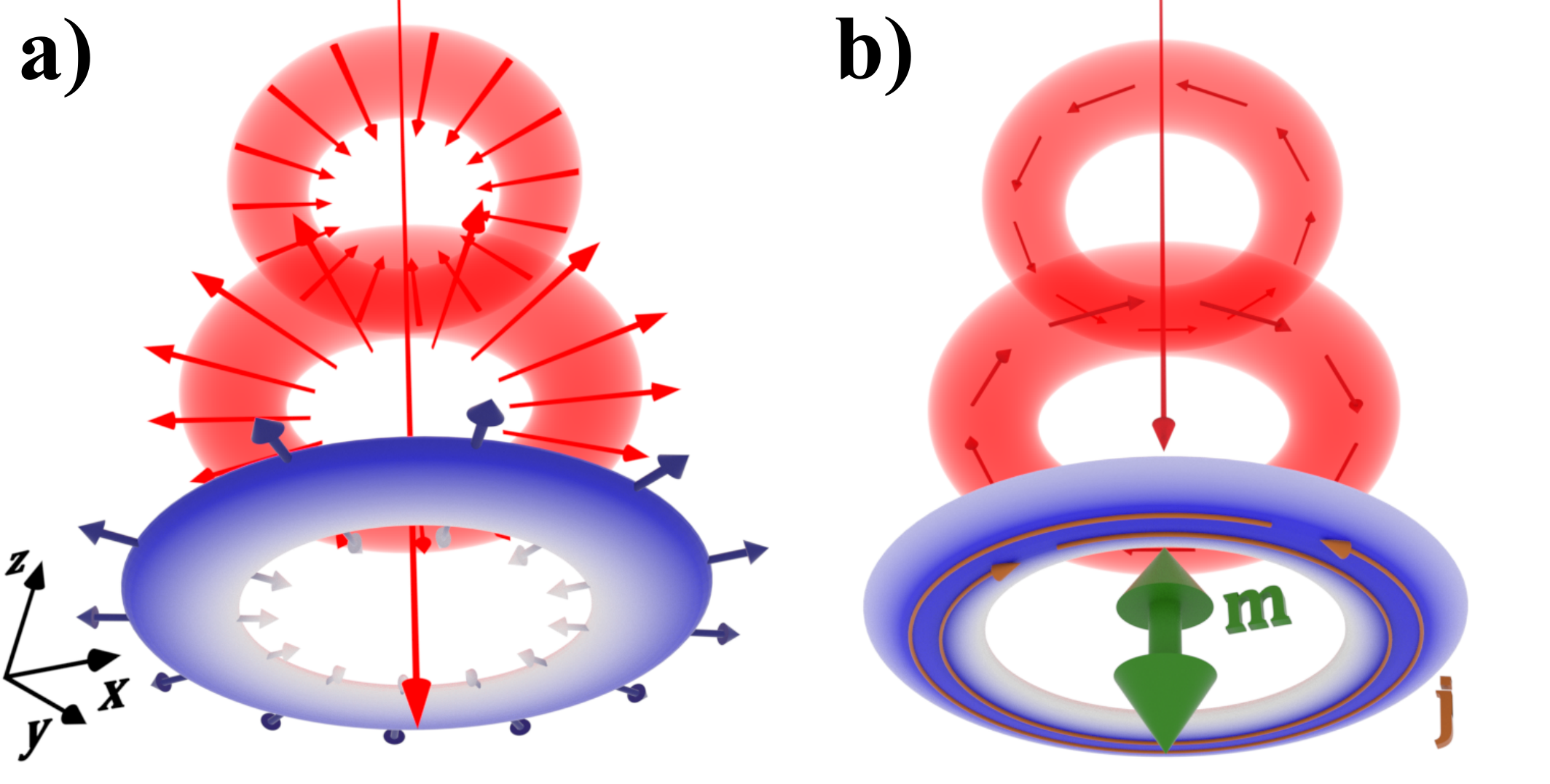}
	\caption{Quantum ring (charge density is marked blue) irradiated by a cylindrical vector
		beam (red rings at two different times). (a) The radially polarized vector beam initiates electric-type transitions
		leading to uniform radially breathing  charge density. (b) The azimuthally polarized vector
		beam generates homogeneous oscillating transient currents (orange arrows) giving rise to
		an oscillating magnetic dipole moment (green arrow). The frequency is set by the driving field.}
	\label{fig1}
\end{figure}

\subsection{Electric Transitions Induced by Radially Polarized  Vector beams }

The interaction of  RVB with quantum rings is illustrated schematically in
Figure~\ref{fig1}a. The coupling to the electric charge is dominant causing
photo-induced transitions, meaning that $\hat{H}^{\rm RVB}_{\rm int}(t)=\hat{H}^{%
	\rm RVB}_{\rm elec}(t)$. Positioning the nano-structure in the plane $z=0$, the
interaction with the RVB associated magnetic field (cf. fields in the SI) reads
$\hat{H}^{\rm RVB}_{\rm magn}(t)=-2\iu\mu_BB'(\pmb{r},t)(z\partial_\rho -\rho%
\partial_z)$. Obviously this has no influence on the confined  electrons in the
$xy$ plane as long as the photon energy $\hbar\omega$ is smaller than the level
spacing of the subbands associated with the confinements in the $z$-direction
(characterized by $U(z)$). RVBs induce electric transitions between states with
an amplitude
\begin{equation}
\begin{split}
M^{\rm RVB}_{\rm int}(t)&=\langle\Psi_{m'}^{s'}|\hat{H}^{\rm RVB}_{\rm el}(t)%
|\Psi_m^s\rangle\\
&=\frac{eA_0\omega }{q_{\perp}}(J_0(q_{\perp}\rho_0)-1)\sin(\omega t) \quad \delta_{m',m}\delta_{s',s}.
\end{split}
\end{equation}
Notably, \textit{no} direct spin-flip transitions are induced by RBVs; and, in
contrast to OAM carrying optical vortex, \textit{no} orbital angular momentum is
transferred to the charge carriers, leading to the selection rule $\Delta m=0$.
Thus, in strictly 1D quantum ring angular momentum and spin states are unaffected
by RVB. For 2D quantum rings $n\to n^\prime$ radial subband transitions are
possible with an amplitude (time-averaged and to a first order in the driving
fields) proportional to the integral
\begin{equation}
\int_0^\infty\du \rho\,R_{n'm}(\rho)R_{nm}(\rho)(J_0(q_{\perp}\rho)-1).
\end{equation}
This  \emph{radial electric  excitation} has a volume character: The evaluated
local dipole moment is the same in all radial directions $\hat{e}_{\rho}$ and
oscillates with a frequency characterized by the energy difference between both
levels $n$ and $n'$. As a result, the \emph{averaged} total moment is zero:
\begin{equation}
\pmb{d}(t)=\int\du \pmb{r}\,\rho_e(\pmb{r},t)\cdot\pmb{r}=0
\end{equation}
where $\rho_e(\pmb{r},t)$ is the (driven) time-dependent charge density.

For detailed and reliable insight, we performed full-numerical space-time-grid
propagation of the three energetically lowest electron states in an irradiated
quantum ring including the external fields to all orders. Figure~\ref{fig2}(a)
displays the resulting charge dynamics induced by the depicted few-cycle external
RVB pulse. The ring radius is $\rho_0=50$ nm and the effective width $\Delta%
\rho=30$ nm. The RVB temporal envelope function $\Omega(t)=\sin[\pi t/T_p]^2$,
where $T_p=2\pi n_p/\omega$ sets the pulse length in terms of the number of optical
cycles $n_p$. We consider a short pulse with $n_p=6$ cycles and a photon energy
$\hbar\omega=8$ meV. The small nano-structure is localized in the low-intense
beam center and away from the first field intensity maximum. Strong multi-photon
processes and the ponderomotive contribution due to $\pmb{A}^2(\pmb{r},t)$ we
found be negligibly small for the light intensity on the ring which was in the
range of $\sim10^4$ W/cm$^2$. As predicted by the analytical treatment, we found
that all the propagated wave functions keep the symmetry in the azimuthal
direction at all times. Field-induced effects are caused by transitions to the
second radial subband leading to charge "breathing" oscillations in radial
direction (we start from the initial states $n=0,\,m=-1,0,1$). The time-dependent
radial expectation value $\langle\rho\rangle$(t) oscillates with a frequency
related to $(E_{n=1,m}-E_{n=0,m})/\hbar$. When the pulse is off, the prodded
charge dynamics goes on due to coherences meaning that every electron state
oscillates between lowest two radial levels.

\subsection{Magnetic Transitions Induced by Azimuthally Polarized Vector Beams}

\begin{figure}[t]
	\includegraphics[width=0.99\columnwidth]{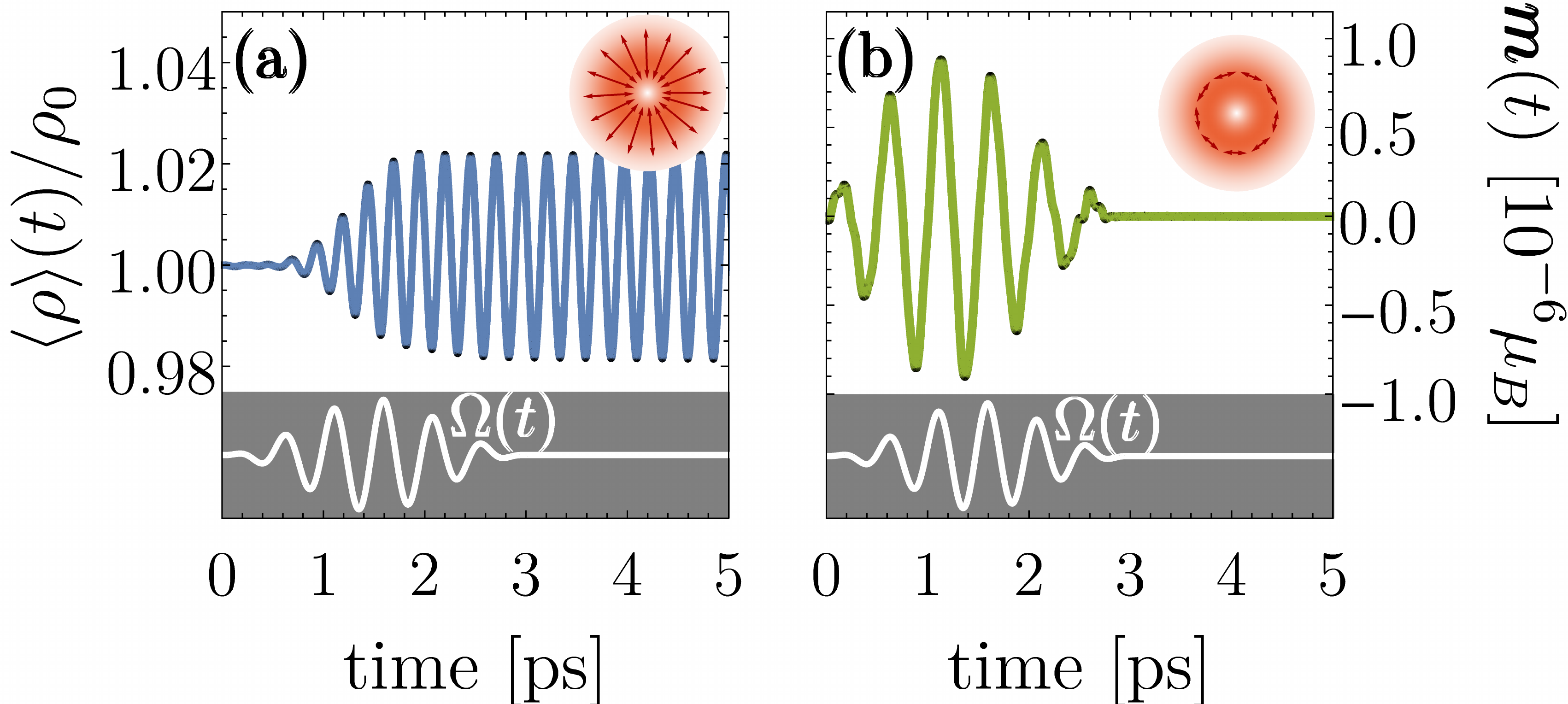}
	\caption{Dynamics from full numerical quantum simulation: (a) time-dependent averaged
		value $\langle\rho\rangle(t)$ of a quantum ring driven by RVB with  six optical cycle duration
		causing  a radial dipole excitation. The oscillation frequency can be
		identified by $(E_{n=1}-E_{n=0})/\hbar$. (b) The time-dependent magnetic moment in
		$z$-direction of  a quantum ring driven by  AVB. The white curves represent the normalized
		electric field amplitude of  the incident light pulses having an intensity  $I\sim10^4$ W
		/cm$^2$ in the area of the ring, and $\hbar\omega=8$ meV.}
	\label{fig2}
\end{figure}

A schematics of an AVB and its action on a quantum ring is shown in Figure~\ref{%
	fig1}(b). It is straightforward to demonstrate that the associated electric
contribution to the light-matter Hamiltonian $\hat{H}_{\rm int}(t)$ vanishes in
the geometry depicted in Figure~\ref{fig1}(b): the electric field is perfectly
azimuthally polarized and therefore $\pmb{r}\cdot\pmb{E}'(\pmb{r},t)\equiv0$.
Thus, the AVB induces \emph{no} electric (dipole) moment.

The light-matter interaction Hamiltonian reduces to $\hat{H}_{\rm int}(t)\equiv%
\hat{H}_{\rm magn}^{\text{\tiny AVB}}(t)$. Note, the contribution to the magnetic
interaction of the type $\hat{e}_z\cdot\hat{\pmb{m}}_B=-\iu\mu_B\partial_\varphi$
does not affect the magnetic quantum number, i.e. the selection rule $\Delta m=0$
is obtained. Generally, the interaction matrix elements have the explicit form
\begin{equation}
\begin{split}
M_{\rm int}^{\text{\tiny AVB}}(t)&=\langle\Psi_{m'}^{s'}|\hat{H}^{\text{%
		\tiny AVB}}_{\rm magn}(t)|\Psi_m^s\rangle\\
&=\mu_B\frac{A_0}{\rho_0}J_{1}(q_{\perp}\rho_0)  \sin(\omega t) \delta_{m',m}
\left[ \left(m + \frac{1}{2} - \frac{1}{2}s\cos(\gamma)\right)
\delta_{s',s}
+ \frac{1}{2}\sin(\gamma)\delta_{s',-s}\right].
\end{split}
\label{eq:int_avb_magn}
\end{equation}
Importantly, in contrast to the RVB, spin-flip can be triggered by AVB (recall
that the eigenstates of $\hat{H}_{0}$ are not eigenstates of $\sigma_z$ nor
$\hat{L}_z$; thus, for instance $\partial_{\varphi}\Psi_{m}^s(\varphi)=\iu(m +%
1/2)\Psi_{m}^s(\varphi) - \frac{\iu s}{2\sqrt{2\pi}}\eu^{\iu(m+1/2)\varphi}\nu^s(%
-\gamma,\varphi)$). Further, the spin-flip transitions are proportional to the
Rashba coefficient $\sin(\gamma)\propto\alpha_R$ (cf. Refs.~\cite{Zhu2008, %
	Quinteiro2011} for EM  homogeneous or OAM pulses).

For $\gamma\rightarrow0$ and starting from equi-populated clock and anti-clockwise
angular momentum quantum numbers $m=-M,-M+1,\dots,M-1,M$ (such as the ground state)
the induced  current density (meaning due to the perturbed states $\delta\Psi_m(t)$)
reads
\begin{equation}
\begin{split}
\pmb{j}(\pmb{r},t)&=\mathbf{ j}_{\rm charge}(\pmb{r},t) + \mathbf{j}_{\rm A}(\pmb{r},t),\\
&=\frac{\hbar}{m_0\rho_0}\sum_{m=-M}^M \left\{\Im\left\{\delta\Psi_m^*%
(\pmb{r},t) \partial_\mathbf{\varphi}\delta\Psi_m(\pmb{r},t)\right\}
- \frac{e}{m_0}\pmb{A}_{\rm AVB}(\pmb{r},t)\left|\delta\Psi_m(\pmb{r},t)\right|^2\right\}, \\
&=\frac{eA_0}{m_0}J_1(q_{\perp}\rho_0)\delta\rho_e\sin(\omega t)\hat{e}_\varphi.
\end{split}
\label{eq:current}
\end{equation}
Obviously,  the $\pm M$ state pair deliver the same (in magnitude) but counter-%
directed current densities and hence $\textbf{j}_{\rm charge}(\pmb{r},t)$ vanishes. Thus,
the current density in the $\varphi$-direction is solely set by the induced charge
density $\delta\rho_e=\sum_{m=-M}^M\left|\delta\Psi_m(\pmb{r},t)\right|^2$ driven
by the vector potential $\pmb{A}_{\rm AVB}(\pmb{r},t)$. It follows that $\delta\rho_e$
oscillates with the frequency of the driving field which means that, beyond transient effects,  no directional
time-averaged current is induced. Since the setup can be tuned to low frequencies,
the effect should be observable. Note, for the AVB the coupling to the electric
field component of light-matter interaction vanishes. Therefore, the resulting
oscillating current caused solely by the time-dependent magnetic vector potential
$\pmb{A}_{\rm AVB}(\pmb{r},t)$  falls thus in the class of a \emph{dynamic
	Aharonov--Bohm effect}.

Since at all time $\pmb{\nabla}\cdot\pmb{j}(\pmb{r},t)\equiv0$, the continuity
equation states that the incident light field does not change the electronic density,
i.e. $\partial_t\rho(\pmb{r},t)=0$ and our system remains locally and at all time
neutral and so does not couple to the electric field part of
AVB. Experimentally, we may sense the action of  AVB by measuring the associated
oscillating magnetic dipole moment
\begin{equation}
\begin{split}
\pmb{m}(t)&=\frac{1}{2}\int\du \pmb{r}\,\pmb{r}\times\pmb{j}(\pmb{r},t)\\
&= \frac{e^2A_0\rho_0}{m_0}J_1(q_{\perp}\rho_0)\delta\rho_e\sin(\omega t)\hat{e}_z.
\end{split}
\end{equation}
In Figure~\ref{fig2}(b) the time-dependent magnetic moment in $z$-direction which
is gathered from a full numerical quantum dynamic simulation is shown. For the
irradiated quantum ring we used the same parameters as for the RVB case. In line
with the analytical predictions eq~\eqref{eq:current}, the build-up and decay
times are locked to the applied external field pointing to a vanishing contribution
$\pm{j}_{\rm charge}(\pmb{r},t)$ to the whole current density (cf. eq~\eqref{eq:current}).
The transient  $\pmb{m}(t)$  vanishes once the pulse is off but one may induce
also an  interference-driven quasi-static component by a combination of two AVB with the frequencies
$\omega$ and  $2\omega$ (not shown here).

The spin-orbit coupling $\hat{H}_{\rm SOI}(t)$ (cf. eq~\eqref{eq:SOI_Bz}) is
mainly determined by the longitudinal component of the magnetic field of the AVB.
The corresponding matrix elements take on the explicit form
\begin{equation}
\begin{split}
M^{\text{\tiny AVB}}_{\rm SOI}(t)&=\langle\Psi_{m'}^{s'}|\hat{H}^{\text{%
		\tiny AVB}}_{\rm SOI}(t)|\Psi_m^s\rangle \\
&=\frac{e\alpha A_0}{\hbar} J_1(q_{\perp}\rho_0)\left[s\sin(\gamma)\delta_{s',s} \right.
\\
&\quad \left. + \cos(\gamma)\delta_{s',-s}\right]
\sin(\omega t)\delta_{m',m}.
\end{split}
\label{eq:int}
\end{equation}
Thus, effectively  AVB results in spin-flip transitions, even to a first order in
the light-matter interaction.  The strength of these transitions is linear in SOI
strength  $\alpha_R$. The matrix element indicates
that even in the presence SOI the AVB does not cause a change in the angular
momentum state. This fact allows  to study pure spin dynamics while the orbital
angular momentum is frozen. We conclude so that a ubiquitous feature of all vector
beam types is that the orbital angular momentum of the electronic states is unaffected.

The further spin-dependent contribution to the AVB-matter interaction is given
by $\hat{H}_{\rm z}(t)$. It describes the direct interaction of the spin state
with the magnetic field component of the vector beam. Generally, it is much weaker
than the spin-orbit interaction, as follows from comparing the prefactors
($e\alpha/\hbar > \mu_B q$). Nonetheless, for completeness we provide an expression
for the matrix elements of this light-matter interaction contribution. Notice, the
magnetic field of the  AVB has also a transverse component which couples to
$\sigma_r$ leading again to spin-flip transitions. In addition to that, the strong
longitudinal field (characterized by $J_0(q_{\perp}\rho)$) gives rise to a
\emph{dynamical} Zeeman effect. The matrix elements can be found analytically
and read explicitly
	\begin{equation}
	\begin{split}
	M_{\rm z}^{\text{\tiny AVB}}(t)&=\langle\Psi_{m'}^{s'}|H^{\rm AVB}_{\rm z}|%
	\Psi_m^s\rangle,\\
	&=\frac{1}{2}g_s\mu_BA_0
	\left[ q_{\parallel}J_{1}(q_{\perp}\rho_0\left(s\sin(\gamma)\delta_{s',s} + \cos(\gamma)%
	\delta_{s',-s}\right)\cos(\omega t)\right. \\
	&\quad\left.+q_{\perp} J_{0}(q_{\perp}\rho_0)\left(s\cos(\gamma)\delta_{s',s} -%
	\sin(\gamma)\delta_{s',-s}\right)\sin(\omega t) \right]
	\delta_{m',m}.
	\end{split}
	\label{eq:int_avb_zee}
	\end{equation}
In practice, both spin-orbit coupling contributions bring about dynamical spin
flip processes while the individual charge currents (associated with the orbital
motion) sum up to zero.

\section{\label{sec:sec3} Atoms Driven by Vector Beams}

\begin{figure*}[t]
	\includegraphics[width=0.95\textwidth]{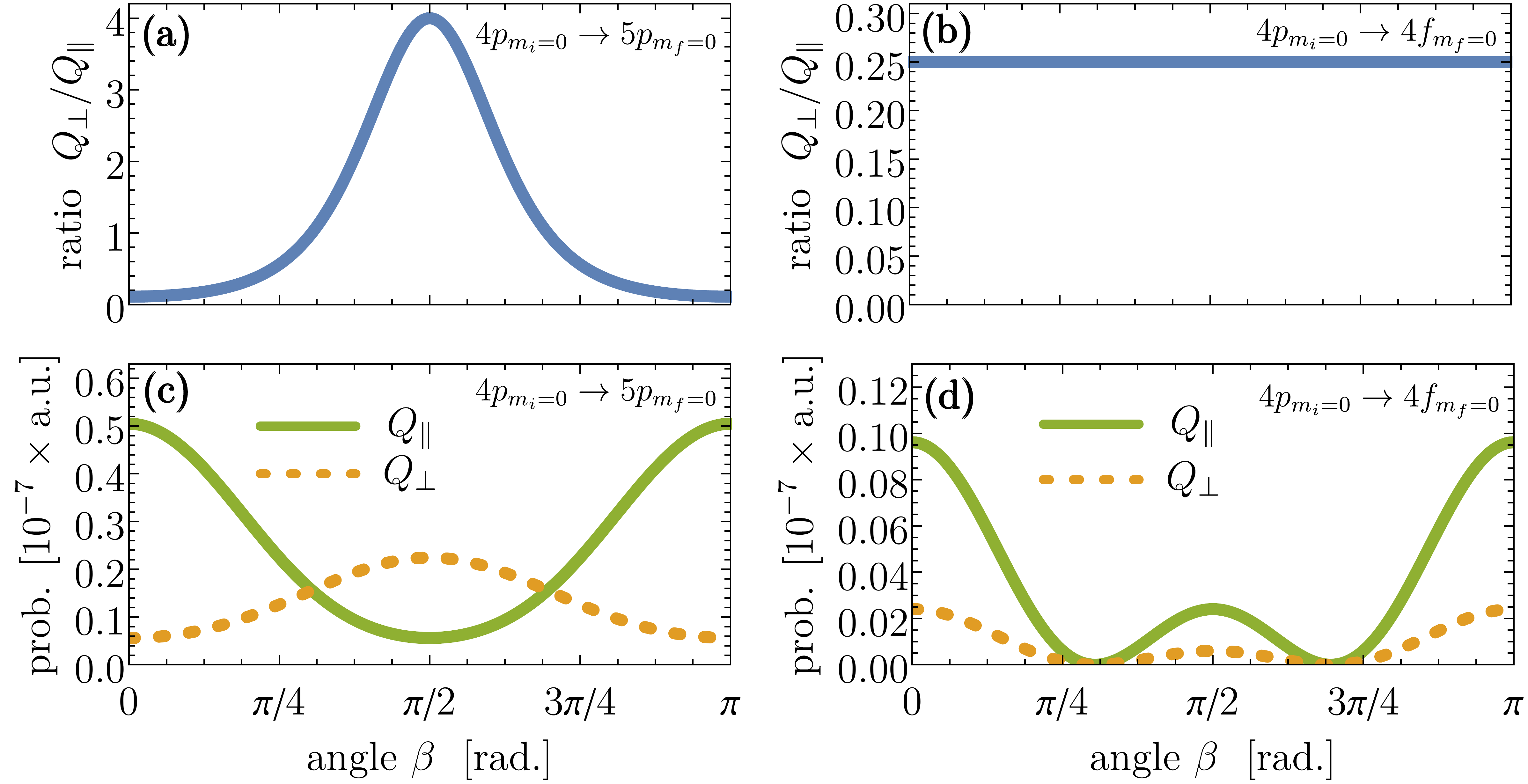}
	\caption{Quadrupole transitions initiated by a vector  laser pulse  with a radial
		polarization. A comparison between the longitudinal and the transverse field
		contributions is shown. Upper row: The ratio between the quadrupole excitation
		probability, labeled by $Q_\perp$ and $Q_\parallel$, for the $p$-$p$ and $p$-$f$
		transitions. Lower row: Explicit quadrupole excitation probabilities for the
		$p$-$p$ and $p$-$f$ transitions. The field has a peak intensity of $I=3.51\times10^{16}$
		W/cm$^2$. The atom is on the optical access and experience a much lower intensity.  The photon energy is  tuned to $\hbar\omega= 4.37$ eV (for $p$-$p$ transitions) or $\hbar\omega= 5.31$ eV ($p$-
		$f$ transitions); all fields have a duration corresponding of thirty optical
		cycles. The static magnetic field is set to $10$ T.}
	\label{fig3}
\end{figure*}

Let us consider as a further case an atomic system in a strong magnetic field
such that SOI is subsidiary compared to $\mu_B B_z(\hat{L}_z + g_s\hat{S}_z)$
(Paschen--Back effect). The electron states with the usual notation $|i\rangle=%
|n_i\ell_i m_is_i\rangle$ are appropriate. The magnetic field sets the quantization
axis ($z$ axis) while the optical axis of the incident VB makes an angle $\beta=%
\measuredangle(\hat{e}_z,q_\parallel)$ with the $z$ axis. We inspect Rydberg
states $|i\rangle$ characterized by the principle, orbital, and magnetic quantum
numbers $n_i$, $\ell_i$ and $m_i$. For photoexcitation of higher Rydberg states,
as already demonstrated for a trapped $^{40}$Ca$^+$ ion by means of OAM vortex
field \cite{schmiegelow2016transfer}, the laser  photon energy is such that the
wave vector $q=\omega/c\ll1$ and thus we expand the oscillating functions in
Eqs.~\eqref{eq:A_RVB} and \eqref{eq:A_AVB} in a Taylor series up to terms of the
first order: $J_1(x)=(1/2)x + \mathcal{O}(x^3) $, $J_0=1+\mathcal{O}(x^2)$ and
$\exp(\iu x)=1+\iu x+\mathcal{O}(x^2)$. In the \emph{local} frame $\left\{\rho',%
\varphi',z'\right\}$ rotated by $\beta$ relative to the $z$ axis, we employ the
rotating wave approximation and obtain for the first-order term of the field of the
RVB
\begin{equation}
\begin{split}
\pmb{A}_{\rm RVB}(\pmb{r}',t) \approx -A_0\left[\frac{1}{2}q_{\perp}\rho'\hat{e}_{\rho'}
+\iu\frac{q_{\perp}}{q_{\parallel}}(1+\iu q_{\parallel}z')\hat{e}_{z'}\right]e^{-\iu\omega t}.
\end{split}
\label{eq:A_RVB_At}
\end{equation}
Interestingly, the quadrupole terms, originating from the longitudinal and the
transverse field distributions have the same prefactors.

Figure~\ref{fig3} shows the results for the photoexcitation process of a trapped
Ca$^+$ ion starting from the initial  $4p_{m_i=0}$ Rydberg state by a \emph{%
	continuous wave} (CW) RVB.
\begin{figure*}[t]
	\includegraphics[width=0.95\textwidth]{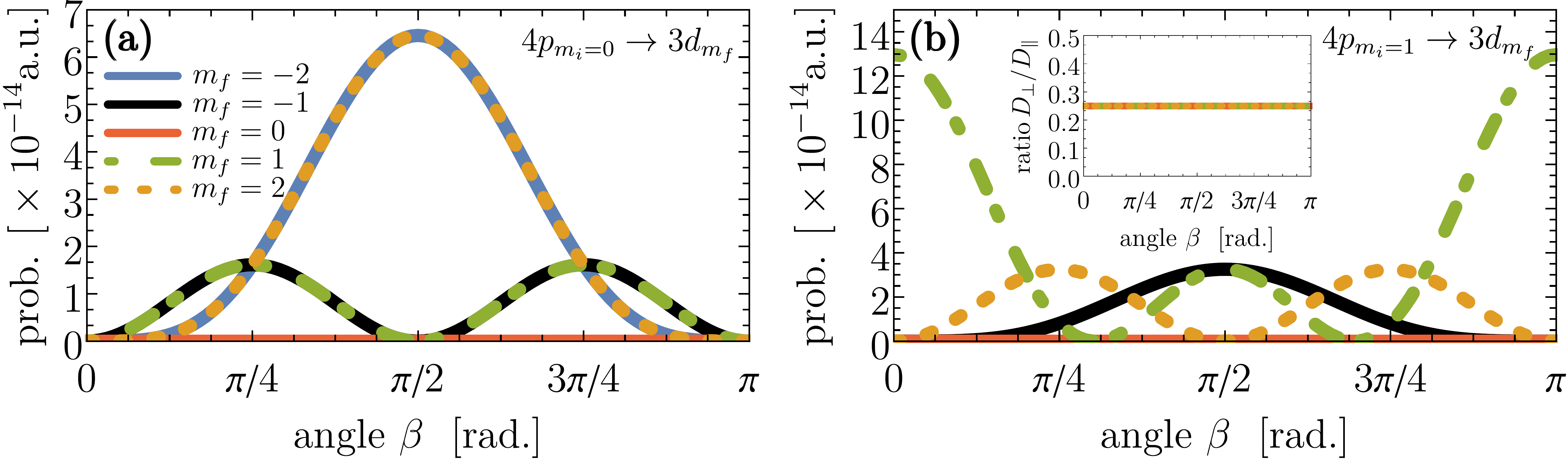}
	\caption{Dipole transitions initiated by an azimuthally polarized vector laser
		pulse. The population of  different magnetic sublevels in dependence on the
		rotation angle $\beta$ between the beam optical axis and the direction of an
		external static magnetic field.  Two initial states $4p_{m_i=0}$ and $4p_{m_i=1}$
		are shown. Inset: The ratio between the dipole transition probabilities originating
		from the longitudinal and the transverse field contributions, labeled by $D_\perp$
		and $D_\parallel$. Similar field parameters where used as for Figure~\ref{fig3}.}
	\label{fig4}
\end{figure*}
The nature of the matter interaction with a radially polarized vector beam is
dominantly electric and is characterized by a strong dipole term stemming from
the (electric)  longitudinal component. However, one can discriminate between
these dipolar and higher order electron transitions by adjusting appropriately
the photon energy to $\hbar\omega=\varepsilon_{5p}-\varepsilon_{4p}$ or $\hbar%
\omega=\varepsilon_{4f}-\varepsilon_{4p}$. The peak amplitude of the vector beam
was chosen to be $A_0=1$ a.u.. In the region of the atom, however, this amplitude
is very small (the prefactor for the transverse electric field is $A_0\omega%
q_\perp$ and for the longitudinal field $A_0\omega\tan\alpha$ with $\alpha=1^\circ$).
The left panels on Figure~\ref{fig3} show the quadrupole transition probabilities
$Q_{\perp}$ and $Q_{\parallel}$, resolved for the longitudinal and transverse field
components, in dependence on the rotation angle $\beta$ for the initial-final
state transition $4p_{m_i=0} \rightarrow5p_{m_f=0}$. Interestingly, for the orbital
momentum conserving quadrupole transition with $\Delta\ell=\Delta m=0$ the ratio
$Q_{\perp}/Q_{\parallel}$ can be steered by rotating the incident vector field
relative to the applied magnetic field (which sets the quantization axis). Parallel
to the magnetic field, the ratio $Q_{\perp}/Q_{\parallel}=1/9$ reveals the
dominating longitudinal component while at an angle of $\beta=90^\circ$ (RVB and
magnetic field are perpendicularly polarized) the transverse component dominates
the photoexcitation process since $Q_{\perp}/Q_{\parallel}=4$. Therefore, in
contrast to a conventional Gaussian mode we can find angular momentum conserving
quadrupole transitions for all possible light field setups due to the special
spatially inhomogeneous character of the vector beam.

The situation changes when exploring the $4p_{m_i=0}\rightarrow4f_{m_f=0}$
quadrupole transition which is characterized by $\Delta\ell=2$. Here, it is not
possible to change the ratio between the longitudinal and transverse field
contributions since $Q_{\perp}/Q_{\parallel}=1/4$ for all rotation angles $\beta$.
Interestingly, we find rotating angles where the quadrupole transitions (either
from the longitudinal or transverse field) vanishes completely. For such a setup
the whole photoexcitation probability of $4p_{m_i=0}\rightarrow 4f_{m_f=0}$
collapses.

For the azimuthal polarization the interaction is fully magnetic since $\pmb{r}%
\cdot\hat{e}_\varphi'=0$ and, therefore, we have no coupling to the electron (note
that $\pmb{r}\cdot\hat{e}_\varphi'$ vanishes for every rotation angle $\beta$).
With same approximations as for the RVB the strong longitudinal field is given (up
to the first order in $r$) by $B_\parallel(\pmb{r}',t)=\iu A_0q_\perp(1+\iu%
q_\parallel z')e^{-\iu\omega t}\hat{e}_z'$ while the transverse field $B_\perp(%
\pmb{r}',t)=\frac{1}{2}A_0q_\perp q_\parallel\rho'e^{-\iu\omega t}{e}_\rho'$. The
homogeneous term in the longitudinal component provides no contribution to the
photo-induced electron transition since it characterizes a monopole interaction.
Consequently, as for the electric type transitions in the case of the RVB the
effective contributions of the longitudinal and transverse field components are
on equal footing but the associated light-matter interaction is dipolar.

In Figure~\ref{fig4} we show the dipole transitions initiated by the spatially
inhomogeneous magnetic field of the AVB for two different initial states $4p_{m_i=0}$
and $4p_{m_f=1}$. As expected, for $\beta=0^\circ$ no photoexcitation processes
can be observed since the magnetic field does not act on the electron states with
zero angular velocity ($m_i=0$). However, as inferred from Figure~\ref{fig4}(a),
for finite rotating angles one can populate final magnetic substates with $m_f\neq0$.
Prominent setups are given by $\beta=\pm45^\circ$ where states with $m_f=\pm1$ and
$m_f=\pm2$ are equally excited while for $\beta=\pm90^\circ$ the final states are
fully characterized by $m_f=\pm2$. Note, that due to the presence of the external
magnetic field, which sets the quantization axis and shifts the energy of the
individual magnetic substates (Zeeman effect), the photon energy of the incident
AVB has to be adjusted to the specific transitions, i.e. $\hbar\omega=\varepsilon_{%
	n_f\ell_f m_f}-\varepsilon_{n_i\ell_i m_i}$.

In Figure~\ref{fig4}(b) the dipolar photoexcitation transitions are depicted for
the initial state $4p_{m_i=1}$. As expected for $\beta=0^\circ$, the only state
which can be excited is characterized by $m_f=1$ since in this case the interaction
between the electron and the light field is angular momentum conserving, i.e.
$\Delta m=0$. Interestingly, at $\beta=\pm90^\circ$ the photoexcitation probability
of $m_f=\pm1$ is the same while at $\beta=\pm45^\circ$ the dominating final state
is characterized by the magnetic quantum number $m_f=2$. Another striking feature,
in  contrast to the RVB, is the ratio between  transverse and  longitudinal
field contributions which is always given by $D_\perp/D_\parallel=0.25$ and thus,
can not be manipulated by the rotation angle $\beta$.

\section{\label{sec:sec4}Summary and outlook}

We explored the nature of the interaction of  atomic and  low dimensional
quantum systems (rings) with EM fields with spatially inhomogeneous polarization
states, called vector beams. In particular, we focused on cylindrical beams with
radial or azimuthal polarization. Although these beams share some common
features with vortex beams carrying orbital angular momentum, like the
intensity profile, their effect on charge carriers is fundamentally different. For
the investigated systems, radially polarized vector beams (RVB) trigger via
electric transitions radial charge oscillations. Azimuthally polarized vector
beams (AVB) generate via a magnetic interaction oscillating magnetic moments.
Despite the presence of the electric field in AVB, it subsumes in a way that it
does not affect the charge. The interaction with AVB is  solely due to the
magnetic vector potential, and can thus be interpreted as a \emph{dynamic}
Aharonov--Bohm effect. In contrast to OAM carrying fields, no unidirectional,
time averaged currents are generated by AVB nor by RVB.
Atomic targets subject to radially polarized light fields show angular momentum
conserving quadrupole transitions which can be manipulated in magnitude by
rotating the field relative to the quantization axis set by an external static
magnetic field. When photoexciting with an azimuthally polarized field, the
special field structure makes it possible to select different magnetic sublevels
(in the final state) by rotating the laser field relative to the quantization axis
of the atomic target.\\


\section{Acknowledgements}

This work was partially supported by the  DFG through SPP1840 and SFB TRR 227.

\begin{appendix}

\section{\label{sec:AppA}}

The electric field of a radially polarized vector beam (RVB) is given by
\begin{subequations}
	\begin{align}
	E_{r}^{\rm RVB}(\pmb{r},t)
	&=A_0\omega_xJ_{1}(q_{\perp}\rho)\sin(q_\parallel z-\omega_xt),\\
	E_{\varphi}^{\rm RVB}(\pmb{r},t)&=0,\\
	E_{z}^{\rm RVB}(\pmb{r},t) & = A_0\omega_x\frac{q_\perp}{q_\parallel}
	J_{0}(q_\perp \rho) \cos(q_\parallel z-\omega_xt),
	\end{align}
	\label{eq:Erad}
\end{subequations}
while the associated magnetic field reads
\begin{subequations}
	\begin{align}
	B_{r}^{\rm RVB}(\pmb{r},t)      &=0,\\
	B_{\varphi}^{\rm
		RVB}(\pmb{r},t)&=A_0\frac{q_\perp^2+q_\parallel^2}{q_\parallel}J_{1}(q_\perp
	\rho)
	\sin(q_\parallel z-\omega_xt),\\
	B_{z}^{\rm RVB}(\pmb{r},t)      &=0.
	\end{align}
	\label{eq:Brad}
\end{subequations}
In the same vein, the electromagnetic fields of the azimuthally polarized vector
beam (AVB) as the sum of two antiparallel Bessel beams read
\begin{subequations}
	\begin{align}
	E_{r}^{\rm AVB}(\pmb{r},t)      &=0,\\
	E_{\varphi}^{\rm
		AVB}(\pmb{r},t)&=-A_0\omega_xJ_{1}(q_\perp \rho)\cos(q_\parallel z-\omega_xt),\\
	E_{z}^{\rm AVB}(\pmb{r},t)      &=0,
	\end{align}
	\label{eq:Eaz}
\end{subequations}
and
\begin{subequations}
	\begin{align}
	B_{r}^{\rm AVB}(\pmb{r},t)      &=A_0q_\parallel J_{1}(q_\perp \rho)
	\cos(q_\parallel z-\omega_xt),\\
	B_{\varphi}^{\rm RVB}(\pmb{r},t)&=0,\\
	B_{z}^{\rm AVB}(\pmb{r},t)      &=-A_0q_\perp J_{0}(q_\perp \rho)\sin(%
	q_\parallel z-\omega_xt),
	\end{align}
	\label{eq:Baz}
\end{subequations}

\section{\label{sec:AppB}}

Using the vector potential in the Poincar\'{e} gauge, i.e. $\pmb{A}'(\pmb{r},t)=%
-\pmb{r}\times\int_0^1\du\lambda\,\lambda\pmb{B}(\lambda\pmb{r},t)$ where the
vector field satisfies $\pmb{r}\cdot\pmb{A}'(\pmb{r},t)\equiv0$, we derive the
expression for the magnetic contribution to the interaction Hamiltonian $\hat{%
	H}_{\rm int}(t)$ from the minimal coupling scheme:
\begin{equation}
H_{\rm magn}(t)=-\frac{e}{2m_0}\left(\pmb{p}\cdot\pmb{A}'(\pmb{r},t) +
\pmb{A}'(\pmb{r},t)\cdot\pmb{p}\right).
\end{equation}
Inserting $\pmb{A}'(\pmb{r},t)$ as well as applying the fundamental identities
$\pmb{p}\cdot\left(\pmb{r}\times\pmb{B}\right) = \left(\pmb{p}\times\pmb{r}\right)%
\cdot\pmb{B}$ as well as $\left(\pmb{r} \times\pmb{B}\right)\cdot\pmb{p} = %
\pmb{B}\cdot\left(\pmb{p}\times\pmb{r}\right)$, we find
\begin{equation}
H_{\rm magn}(t)=\frac{e}{2m_0}\left(\pmb{p}\times\pmb{r}\right) \cdot
\int_0^1\du\lambda\,\lambda\pmb{B}(\lambda\pmb{r},t)
+ \frac{e}{2m_0}\int_0^1\du\lambda\,\lambda\pmb{B}(\lambda\pmb{r},t)\cdot
\left(\pmb{p}\times\pmb{r}\right).
\label{eq:HmanGen}
\end{equation}
From elementary quantum mechanical algebra we know that
\begin{equation}
\left(\pmb{p}\times\pmb{r}\right)\cdot\pmb{B} = \pmb{B}\cdot\left(\pmb{p}\times%
\pmb{r}\right) -  \left[\pmb{B},\left(\pmb{p}\times\pmb{r}\right)\right]_{-}.
\label{eq:comm}
\end{equation}
Using $\pmb{p}\times\pmb{r}=-\pmb{r}\times\pmb{p}$ and we can find the commutator
\begin{equation}
\begin{split}
\left[\pmb{B},\left(\pmb{p}\times\pmb{r}\right)\right]_{-} &=
\left[\left(\pmb{r}\times\pmb{p}\right),\pmb{B}\right]_{-} \\
&=\varepsilon_{ijk}\left[x_jp_k,B_i\right]_{-} \\
&=\varepsilon_{ijk}x_j\left[p_k,B_i\right]_{-} + \varepsilon_{ijk}%
\underbrace{\left[x_j,B_i\right]_{-}}_{=0}p_k \\
&=\pmb{r}\cdot(\pmb{p}\times\pmb{B}).
\end{split}
\end{equation}
By using now Amp\`{e}re--Maxwell law~\cite{jackson1975electrodynamics} the
commutator can be reformulated further:
\begin{equation}
\begin{split}
\left[\pmb{B},\left(\pmb{p}\times\pmb{r}\right)\right]_{-} &=\pmb{r}\cdot(%
\pmb{p}\times\pmb{B})\\
&=-\iu\hbar\pmb{r}\cdot\left[\pmb{\nabla}\times\pmb{B}(\pmb{r},t)\right]\\
&=-\frac{\iu\hbar}{c^2}\pmb{r}\cdot\frac{\partial\pmb{E}(\pmb{r},t)}{\partial t}.
\end{split}
\end{equation}
Furthermore, by assuming a harmonic wave we find that $\partial_t\pmb{E}(%
\pmb{r},t)\sim -\omega_x\pmb{E}(\pmb{r},t)$ and obtain the final expression for
Hamiltonian containing the commutator
\begin{equation}
\begin{split}
H_{\rm magn}^{\rm comm.}&=-\int_0^1\du \lambda\,\lambda\frac{e}{2m} \left[\pmb{B},%
\left(\pmb{p}\times\pmb{r}\right)\right]_{-}\\
&=\frac{\iu e}{2}\frac{\hbar\omega_x}{m_0c^2}\int_0^1\du\lambda\,\lambda\pmb{r} \cdot%
\pmb{E}(\pmb{r},t),
\end{split}
\end{equation}
which can be safely neglected by noticing that the prefactor $\hbar\omega_x/m_0
c^2 < 10^{-4}$ even for photon energies in the (X)UV regime. Furthermore, in the
case of an AVB $\pmb{r} \cdot \pmb{E}(\pmb{r},t)\equiv0$ (azimuthal polarization).
As a consequence, the magnetic part of the interaction Hamiltonian is
\begin{equation}
\begin{split}
H_{\rm magn}(t)&=\frac{e}{m_0}\left[-\int_0^1\du\lambda\,\lambda\pmb{B}(\lambda%
\pmb{r},t)\right] \cdot \left(\pmb{r}\times\hat{\pmb{p}}\right)
\\
&=2\pmb{B}'(\pmb{r},t)\cdot\widehat{\pmb{m}}_B
\end{split}
\end{equation}
where $\pmb{B}'(\pmb{r},t)=-\int_0^1\du\lambda\,\lambda\pmb{B}(\lambda\pmb{r},t)$
and the magnetic moment operator is $\widehat{\pmb{m}}_B=(e/2m)\,\pmb{r}\times%
\hat{\pmb{p}}$.

\end{appendix}


%

\end{document}